\documentclass[useAMS,usenatbib]{mn2e}
\pdfoutput=1
\usepackage{graphicx,color}

\usepackage{bm}
\usepackage{natbib}
\usepackage{amsmath}
\usepackage{xspace}

\def\sigv{\langle \sigma v\rangle}
\def\Om{\Omega_{\rm M}}
\def\Om0{\Omega_{\rm M,0}}

\def\Ob0{\Omega_{\rm B,0}}
\def\OQ{\Omega_{\rm Q}}
\def\OQ0{\Omega_{\rm Q,0}}

\newcommand{\be}{\begin{equation}}
\newcommand{\ee}{\end{equation}}

\newcommand{\beq}{\begin{equation}}
\newcommand{\eeq}{\end{equation}}
\newcommand{\bdm}{\begin{displaymath}}
\newcommand{\edm}{\end{displaymath}}
\newcommand{\bea}{\begin{eqnarray}}
\newcommand{\eea}{\end{eqnarray}}
\newcommand{\bt}{\begin{tabular}}
\newcommand{\et}{\end{tabular}}

\def\la{\langle}

\def\gio{GBSC}

\def\eg{{\em e.g.}\xspace}
\def\ie{{\em i.e.}\xspace}

\def\Ms{\, h^{-1} \, {\rm M}_{\odot}}

\def\kMpc{\, h \, {\rm Mpc}^{-1}}

\def\la{\langle}

\newcommand{\lsi}{\,\raisebox{-0.13cm}{$\stackrel{\textstyle<}
{\textstyle\sim}$}\,}

\title[Extragal. $\gamma$-rays from DM: a PS computation]{Extragalactic gamma-ray signal from Dark Matter annihilation: a power spectrum based computation}
\author[P.~D.~Serpico, E.~Sefusatti, M. Gustafsson, and G.~Zaharijas]{P. D. Serpico$^{1}$\thanks{E-mail:
serpico@lapp.in2p3.fr (PDS)}, E. Sefusatti$^{2,3}$\thanks{E-mail:
esefusat@ictp.it (ES)}, 
M. Gustafsson$^{4,5}$\thanks{E-mail: mgustafs@ulb.ac.be (MG)}, and G. Zaharijas$^{2,3}$\thanks{E-mail: gzaharij@ictp.it (GZ)}\\
$^{1}$LAPTh, Univ. de Savoie, CNRS, B.P.110, Annecy-le-Vieux F-74941, France\\
$^{2}$Institut de Physique Th\'eorique, CEA/DSM/IPhT, Unit\'e de recherche associ\'ee au CNRS, CEA/Saclay 91191 Gif-sur-Yvette France\\
$^{3}$Abdus Salam International Centre for Theoretical Physics, Strada Costiera 11, 34151, Trieste, Italy\\ 
$^{4}$Dipartimento di Fisica ``Galileo Galilei'' Universit\`a di Padova, Via Marzolo 8, 35131 Padova, Italy\\
$^{5}$Service de Physique Th{\'e}orique, Universit{\'e} Libre de Bruxelles, CP225, Bld du Triomphe, 1050 Brussels, Belgium}

\begin{document}

%\date{Accepted ???. Received ???; in original form ???}

\pagerange{\pageref{firstpage}--\pageref{lastpage}} \pubyear{????}

\maketitle

\label{firstpage}

\begin{abstract}
We revisit the computation of the extragalactic gamma-ray signal from cosmological dark matter annihilations.  The prediction of this signal is notoriously model dependent, due to different descriptions of the clumpiness of the dark matter distribution at small scales, responsible for  an enhancement with respect to the smoothly distributed case. We show how a direct computation of this ``flux multiplier'' in terms of the nonlinear power spectrum offers a conceptually simpler approach and may ease some problems, such as the extrapolation issue. 
In fact very simple analytical recipes to construct the power spectrum yield results similar to the popular Halo Model expectations, with a straightforward alternative estimate of errors. For this specific application, one also obviates to the need of identifying (often literature-dependent) concepts entering the Halo Model, to compare different simulations. 
\end{abstract}


\section{Introduction}
To unveil the particle physics nature of the Dark Matter (DM) inferred by astrophysical and cosmological observations, a plethora of  strategies are currently pursued, including production in high energy colliders and direct detection of recoils in underground detectors \citep[see the monograph in][for a review]{Bertone2010}. However, indirect detection of Galactic and extragalactic by-products of DM annihilations (or decays) is the sole way to access information on the DM {\it remotely}, \ie in the environments where evidence for its presence has been collected.  In particular, with the DM in the form of Weakly Interacting Particles (WIMPs) signals in gamma-rays are expected from the inner halo or center of the Milky Way or from Milky Way halo substructures (either known via their baryonic counterparts, dwarf galaxies, or as ``dark satellites''). All these signals, while promising,  depend on the ``local'' (in cosmological terms) DM environment, related for example to the assembly history of our Galaxy.  On the other hand, at high Galactic latitudes the diffuse signal from the Milky Way halo is expected to be roughly comparable with the diffuse extragalactic one. This Extragalactic DM annihilation Flux (EDMF) depends in a {\em statistical sense} on how DM is distributed at different scales, and how this distribution evolves with redshift.  Although typically more challenging to detect than its Galactic counterpart, the EDMF retains important cosmological information otherwise impossible to access. Since the underlying particle physics parameters are the same, one might hope to corroborate an eventual signal of DM annihilation by looking at this high-latitude flux, or to isolate the cosmological information by comparison with the Galactic signal. 
 
The purpose of this article is to revisit the computation of the ``flux multiplier'', which is the main poorly known ingredient relating the EDMF to the parameters determining the Galactic flux. Although this issue has been subject to a lot of attention in different contexts \citep[see e.g.][]{BergstromEtal2001,UllioEtal2002,SilkTaylor2003} here we advocate a more straightforward computation via direct integration of the nonlinear power spectrum (PS). Although formally equivalent to the Halo Model (HM) framework in configuration space, especially when discussing individual component uncertainties and cosmology-dependency, this approach has the great advantage of involving mostly model-independent quantities (i.e. whose definitions do not rely on the specific parameterization of the variables and quantities entering the HM) which can be extracted directly from simulations. In addition, some uncertainties on clustering properties are easier to assess in the this context.
Even when a HM approach is used to guide the extrapolation to the very small scales,  the PS approach has the benefit to isolate the ``effective combination of parameters'' responsible for the signal.

This article is structured as follows. In Sec.~\ref{formalism} we introduce the central notion of {\it flux multiplier} $\zeta(z)$ and recap the main formulae used for its computation within the HM. In Sec.~\ref{pscomputation} we show how the ``systematic'' uncertainty in the signal can be more effectively reformulated in terms of the PS extrapolation.  Finally, in Sec.~\ref{conclusions} we discuss our results and comment on possible future directions. 

%%%%%
%%%%%%%%%%%%%%%%%%%%%%%%%%%%%%%%%%%%%%%%%%%%%%%%%%%%%%%%%%%%%%%%%%%%%%%%%%%%%%%%%%%%%%%%%%%%%
\section{Standard Formalism}
\label{formalism}
%%%%%%%%%%%%%%%%%%%%%%%%%%%%%%%%%%%%%%%%%%%%%%%%%%%%%%%%%%%%%%%%%%%%%%%%%%%%%%%%%%%%%%%%%%%%%

Formally, for a constant annihilation cross section $\sigv$ the flux (number of photons per energy interval, unit area, time and solid angle) from DM particles with mass $m_\chi$ can be written \citep[see \eg][]{AndoKomatsu2006}
\be
\phi(E,\hat{\Omega}) = \frac{\sigv\,c}{8\,\pi\,m_\chi^2}\int dz \frac{e^{-\tau(z,E)}}{H(z)}\frac{\rho^2(z,\hat{\Omega})}{(1+z)^{3} }\frac{dN(E',z)}{dE'}\,,\label{genereq}
\ee
with $H(z) $ the Hubble expansion function, $\rho$ referring to the DM density at redshift $z$ in the angular direction $\hat \Omega$,  $e^{-\tau(z,E)}$ accounting for 
absorption onto the extragalactic background light, and $dN/dE$ being the emitted spectrum per annihilation, with $E'$ denoting the energy corresponding to present-day value $E$ at redshift $z$. In general this spectrum includes not only the prompt photons, from lines or $\pi^0$ decays, but also secondary emission from energy losses of other particles (such as inverse-compton from $e^{\pm}$ onto the CMB). Equation~(\ref{genereq}) is usually rewritten isolating the directional dependence contained in the density contrast field $\delta$ as: 
\be
\rho(z,\hat{\Omega})\!\equiv\! [1\!+\!\delta(z,\hat{\Omega})]\,\bar{\rho}(z)\!=\! [1\!+\!\delta(z,\hat{\Omega})]\Omega_{\rm DM}\,\rho_c (1\!+\!z)^3\,.
\ee
Due to the deeply nonlinear regime of matter perturbations nowadays, a very good approximation is to neglect the ``average'' matter contribution in the equation above (the ``1'' term in square brackets) and concentrate on the one due to the clumpiness (the $\delta$ term).
In addition, if, as here, the angular dependence is not of interest and all one cares about is the average flux over the angular direction, one  has
\be
\phi(E)=\frac{c\,\sigv(\Omega_{\rm DM}\rho_c)^2}{8\pi\,m_\chi^2}\!\int\! dz \frac{e^{-\tau}(1+z)^3}{H}\,\zeta(z)\,\frac{dN}{dE'}\,,\label{finaleq}
\ee 
where we defined the {\it flux multiplier}
\be
\zeta(z)\equiv\langle \delta^2(z,\hat{\Omega})\rangle\,,
\label{eq:zetaDef}
\ee
namely the variance of density fluctuations over the sky at a given epoch. The largest uncertainty in the EDMF computation stems from $\zeta(z)$, on which we concentrate henceforth. 

In order to compute the above quantity, one customarily resorts to the Halo Model (HM) framework \citep[see][for a review]{CooraySheth2002}.
 The HM assumes that all the mass in the Universe is contained in virialized objects ({\em halos}) fully characterized by their mass. As a consequence, the statistical properties of the mass density field are determined by the spatial distribution of matter within an individual halo {\em and} by the spatial distribution of halos, assumed not to overlap one with respect to the other. For this calculation the crucial quantities are given by the number density of halos of a given mass ({\em halo mass function}) and by the density distribution of each halo ({\em halo density profile} and {\em concentration}).
This allows one to write $\zeta(z)$ as in~\citet{UllioEtal2002}
\be
\zeta(z)= \frac{1}{\Omega_{\rm M}\rho_c}\int_{M_{\rm min}} d M  \frac{d n}{dM} M\frac{\Delta_v(z)}{3} \langle F\rangle\,,\label{zetaz}
\ee
with  $dn/dM$ the {\it comoving} density of halos per unit mass, $\Delta_v$ the mean halo over density, and $F$ being the function 
\be
F(M,z)\equiv c_v^3(M,z)\frac{\int_0^{c_v}dx\,x^2\kappa^2(x)}{\left[\int_0^{c_v} dx\,x^2\,\kappa(x)\right]^{2}}\,,\label{FMz}
\ee
which depends on the DM {\em halo} density $\rho=K(M,z)\kappa(x)$, where $K$ includes the cosmology dependence in terms of the variables $\{M,z\}$ and $\kappa(x)$ is the assumed universal shape function determined by numerical simulations.
 Also, one assumes that the halo density is non-vanishing only within a radius $R_v$ ({\it virial radius}), which is conventionally parametrized via the {\it concentration parameter} $c_v=R_v/r_s$; the dimensionless variable $x$ is just $r/r_s$. The halo properties need not to be universal: in that case $F$ has to be intended as an average over the probability distribution of the relevant parameters (most notably $c_v$). Finally, $\frac{d n}{dM}$ is commonly normalized by imposing that all mass resides in some halo.

In general, a more faithful description of simulations requires accounting for halo ``sub-structures'' (sub-halos) which in turn have their own mass-function, concentration and shape/profile properties.
Also, one can distinguish different ``halo populations'', according to the degree of dynamical interaction they undergo with other structures: the ones the HM is typically compared to are the so-called ``distinct'' or ``isolated'' halos. All in all, in order to perform an estimate of $\zeta (z)$, the usual practice is to {\it fit} from simulations the following quantities: $\frac{d n}{dM}(M,z)$, $\Delta_v(z)$, $c_v(M,z)$ (as well as its distribution around the mean), $\kappa(x)$ (as well as its distribution around the mean
and possible $z-$evolution), and similar quantities for each different category of objects: sub-halos, non-distinct halos, etc. Additionally, some approximations (\eg spherical
shape of the halos) are implicitly assumed.

%%%%%%%%%%%%%%%%%%%%%%%%%%%%%%%%%%%%%%%%%%%%%%%%%%%%%%%%%%%%%%%%%%%%%%%%%%%%%%%%%%%%%%%%%%%%%
\section{Direct computation of $\zeta$ from the power spectrum}
\label{pscomputation}
%%%%%%%%%%%%%%%%%%%%%%%%%%%%%%%%%%%%%%%%%%%%%%%%%%%%%%%%%%%%%%%%%%%%%%%%%%%%%%%%%%%%%%%%%%%%%
\begin{figure*}
{\includegraphics[width=0.45\textwidth]{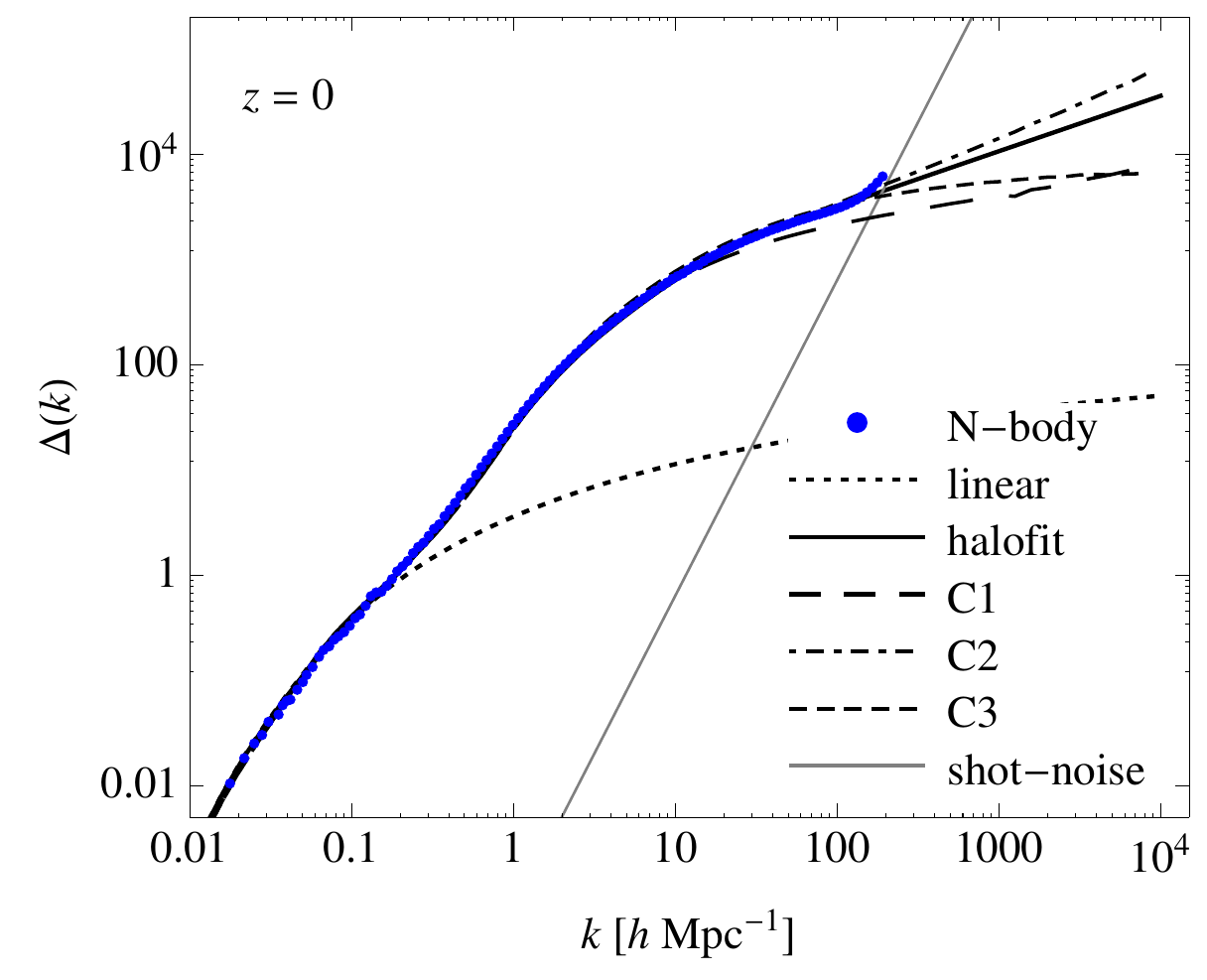}}
{\includegraphics[width=0.45\textwidth]{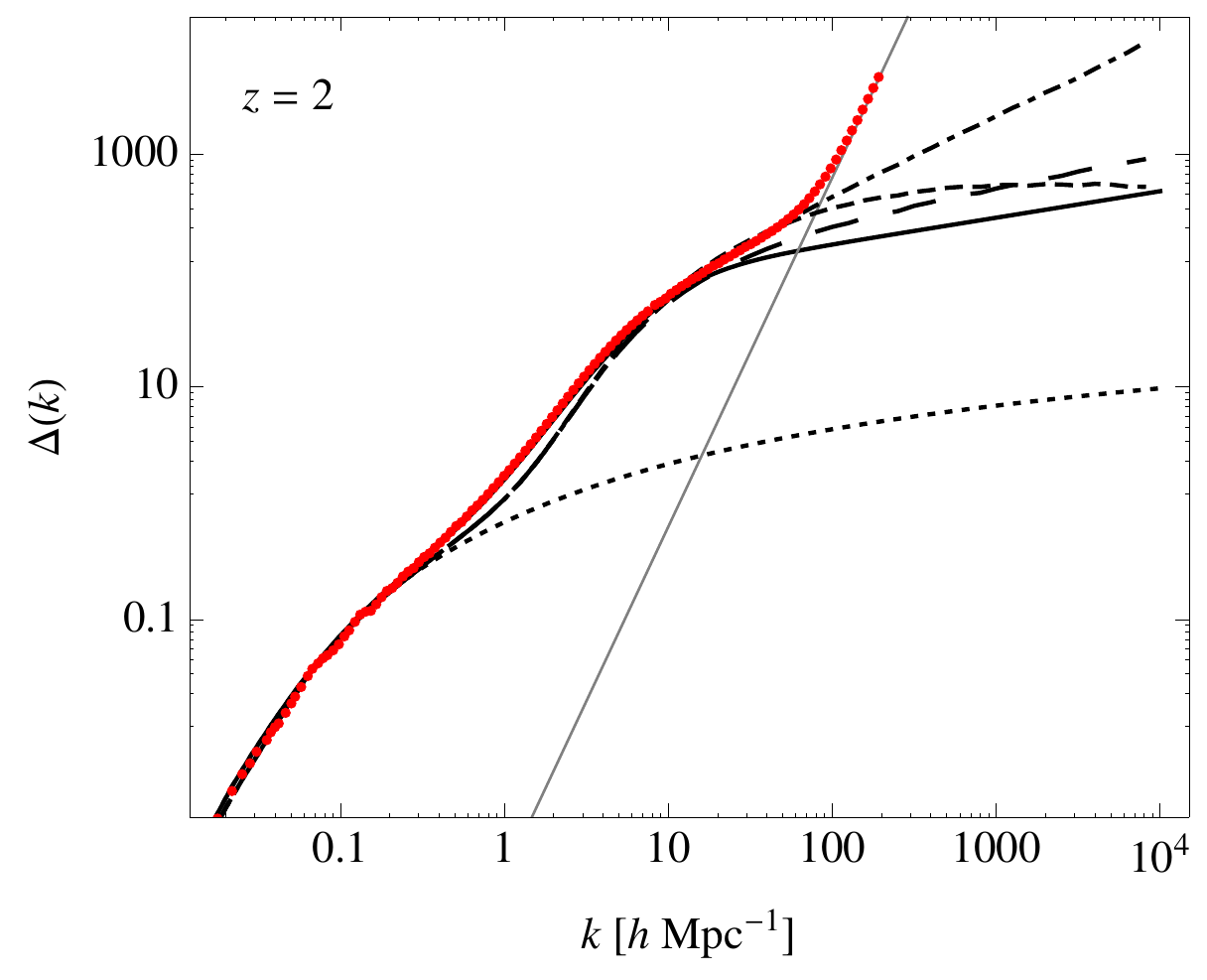}}
\caption{Dimensionless power spectrum $\Delta(k)$ ({\em data points}) from the Millennium Simulation~\citep{SpringelEtal2005} at $z=0$ ({\it left panel, blue}) and at $z=2$ ({\it right panel, red}).
The dotted curve denotes the corresponding linear power spectrum, the continuous curve the \texttt{halofit}~\citep{SmithEtal2003} prediction while the three different HM predictions of~\gio\, are given by the long-dashed curves for the $C1$ model, the dot-dashed for $C2$ and the short-dashed for $C3$. The gray, straight line represent the the shot-noise limit expected if the simulation particles were a Poisson sampling from a smooth underlying density field. }
\label{fig:DeltaNL}
\end{figure*}

While the HM has proven extremely useful for the semi-analytical modeling of nonlinear clustering,  the EDMF does not really depend directly on the many different parameters and functions of the HM, but {\it only} on the PS at very small scales. In fact, the flux multiplier of Eq.~(\ref{eq:zetaDef}) can also be written as the $r\to 0$ limit of the two-point correlation function (2PCF) $\equiv\langle \delta\left({\vec x}+{\vec r} \right) \delta \left( {\vec x} \right)\rangle$. Perhaps this is best seen if one thinks of fluctuations of the DM field in {\em momentum space} and rewrites $\zeta(z)$ in terms of the {\it nonlinear} matter power spectrum $P_{NL}$---defined as a Fourier transform of the 2PCF---as 
\be
\zeta(z) \equiv  \lim_{r \to 0} \int^{k_{max}}\frac{d\,k}{k} \frac{\sin{kr}}{kr}\Delta_{NL}(k,z)
\,.\label{zetazNL}
\ee
Here $\Delta_{NL}(k)$ is the dimensionless nonlinear PS defined as~\footnote{Note that our convention for $P_{NL}(k)$ is linked to the $P_{NL}^{B}(k)$ defined e.g. in~\citep{BernardeauEtal2002} via $P_{NL}(k)=4\,\pi P_{NL}^{B}(k)/(2\pi)^3$; the $\Delta_{NL}(k)$'s are instead the same and correspond to standard definitions.} $\Delta_{NL}(k)\equiv k^3P_{NL}(k)/(2\pi^2)$, while the effective ultraviolet (UV) cutoff $k_{max}$ corresponds to the acoustic oscillation or free-streaming damping scale $k_{cut}$, dependent on the DM candidate's kinetic decoupling temperature from the plasma in the early Universe. Typical expectations for WIMPs are in the range $k_{cut}\in [10^{5}, 10^{8}]\kMpc$ at the end of the linear regime of structure formation \citep[see][and refs. therein]{GreenHofmannSchwarz2005, Bringmann2009}, which  are usually mapped into corresponding  ``smallest halo'' masses of the order of $M_{min}\sim 4\pi/3 (\pi/k_{\rm max})^3\in [10^{-12}, 10^{-3}]\,M_\odot$~\footnote{In the HM, UV cutoffs enter both via the $r\to 0$ limit of the halo profile (HP)
and via the minimum halo mass. When extrapolating HPs found in current numerical simulations one finds that  the inner halo (e.g. $r\ll$ scale radius $r_s$ in the popular NFW HP~\citep{NFW}) contributes only modestly to the global annihilation
signal; thus $M_{min}$ de facto regulates the extrapolation uncertainty. Also note that while the mapping between a cutoff in $k$-space and $M_{min}$ has some ambiguities that are intrinsic to the HM, the
associated error is subleading compared to the extrapolation error.}. In the following, we shall consider $k_{max}=10^{6}\kMpc$ as a representative value. Notice that the low-$k$ region contributes quite insignificantly to the above integral, since the fluctuations of the field are still in the perturbative regime. The size of the observable Universe provides the physical infrared cutoff, $k_{\rm min}\sim \cal{O}$(10$^{-3})\kMpc$.

Unfortunately, there is no rigorous theoretical way to compute $\Delta_{NL}(k,z)$. Linear theory  can be extended in perturbation theory only to the mildly nonlinear regime ($k\lsi 0.3\kMpc$ at $z=0$) where the matter density field can be treated as a fluid in the single-stream approximation \citep[see][for a review]{BernardeauEtal2002}. At smaller scales, the occurrence of shell-crossing and the eventual virialization of over-dense regions prevents the problem to be treated rigorously from first principles, and simulations are necessary for quantitative inferences.
Yet, theoretical considerations based on the theory of spherical collapse and stable clustering, that is the idea that virialized regions decouple from Hubble expansion keeping a constant physical size, allowed \citet{HamiltonEtal1991} to derive a mapping between the linear and the nonlinear matter correlation function, eventually extended to the PS in \citet{PeacockDodds1994, PeacockDodds1996}.  Similarly, the popular \texttt{halofit} formula \citep{SmithEtal2003} provides a cosmology-dependent mapping (fitted to N-body simulations) from the linear to nonlinear PS, significantly improving the accuracy of the earlier approaches. 

Of course, one can also extrapolate the nonlinear PS predicted within the HM approach.
In this framework, the PS can be decomposed into the sum of a term coming from the correlation of two DM particles within the same halo ({\it 1-halo term}) and a term describing the correlation between particles belonging to distinct halos ({\it 2-halo term}), related to the two-point function of the halo distribution. The highly nonlinear regime is clearly described by the 1-halo contribution which dominates the enhancement factor $\zeta$. Notice that the 1-halo contribution depends on the same ingredients required by the usual calculation of the annihilation flux described in the previous section, just expressed in $k$-space.
If the HM could reproduce completely the PS extracted from simulations, going through the HM or a direct computation via integration of the PS would produce
identical results.  
 However, it is known that several aspects are missed in the HM, especially in describing deeply nonlinear scales. 
For instance \citet[][hereafter \gio]{GiocoliEtal2010B} extends the basic picture to include the effect of a subhalo population and the scatter of the concentration parameter.

\begin{table*} 
\renewcommand{\arraystretch}{1.1} \renewcommand{\tabcolsep}{0.26cm}
\begin{minipage}{168mm}
\label{tabzeta}
\begin{tabular}{||c||c||c|c|c|c||c||c|c|c|c||c||c||c||}
 & \multicolumn{13}{c||}{Enhancement factor $\zeta\times 10^{-3}$} \\
\cline{2-14}
 & \multicolumn{5}{|c||}{$k_{\rm max}=50\kMpc$} & \multicolumn{5}{c||}{$k_{\rm max}=10^4 \kMpc$}   & \multicolumn{3}{c||}{$k_{\rm max}= 10^6\kMpc$} \\ 
 \cline{2-14}
$z$ & MS   & HF   &  C1   &  C2   &  C3   & MS ext. & HF  &  C1   &  C2   &  C3   & MS ext. & MSII ext.
& HF \\
\hline\hline
0 & 2.7  & 2.6  & 2.4   & 2.6   & 3.1   & 11/310      & 66  & 24    & 100   & 29    & 18/31000    & 46/7700  & 760  \\
\hline
1 & 0.65 & 0.52 & 0.53  & 0.65  & 0.72  & 1.8/110     & 5.4 & 5.5   & 27    & 6.1   & 2.9/11000   & 8.8/3300  & 25 \\
\hline  
2 & 0.24 & 0.20 & 0.21  & 0.26  & 0.27  & 0.45/74    & 1.7 & 2.8   & 15    & 2.8   & 0.71/7500    & 3.3/1300 & 5.6 \\
\end{tabular}
\caption{Enhancement factor $\zeta$ (divided by 10$^3$) at four different redshifts from the Millennium Simulation (MS) power spectrum and its extrapolations, the \texttt{halofit} formula (HF), and three models of~\gio\,, denoted C1, C2, C3 as in the original work. Results from the extrapolations in Eq.s~(\ref{extrap0}) and (\ref{extrap1}) are given in the form ``min''/``max'', respectively . Columns 2 to 6 assume an integration up to $k_{\rm max}=50\kMpc$; columns (7 to 11) assume $k_{\rm max}=10^{4}\kMpc$ while columns 12 and 13 present the result of extending the integration up to $k_{\rm max}=10^{6}\kMpc$. Column 13 is like column 12, but for the Millennium-II Simulations. }
 \end{minipage} \end{table*}

It is thus useful to compute $\zeta(z)$  by performing the integral in Eq.~(\ref{zetazNL}), relying as much as possible directly on the results of simulations. 
A typical result for the nonlinear matter PS measured in simulations at $z=0$ and at $z=2$, is reproduced by the data points in Fig.~\ref{fig:DeltaNL}, adapted from~\citet{SpringelEtal2005}. This is compared with the linear PS prediction for the same cosmology ({\em dotted curves}) and the nonlinear predictions of the \texttt{halofit} fitting formula ({\em continuous curves}). Table~\ref{tabzeta} shows the enhancement factor $\zeta$ computed using the \texttt{halofit} prediction and assuming a sharp cut-off in the integral given by $k_{max}=50$, $10^4$ and $10^6\Ms$ in the columns 3, 8, and 13, respectively and for $z=0$, $1$ and $2$. For instance, we obtain $\zeta=2.7\times 10^3$ for $k_{max}=50\kMpc$ at $z=0$, very close to the simulation results\footnote{In our calculation we add to the \texttt{halofit} prescription the correction, relevant at small scales, provided by J. Peacock here: http://www.roe.ac.uk/$\sim$jap/haloes/}.
Lacking a firm theory, we can tentatively assume the above extrapolations as ``fiducial'' prediction for the PS.  
We also show three other HM predictions corresponding to different mass-concentration relations and sub-halo models,  denoted $C1$, $C2$ and $C3$ as in~\gio\,, with the long-dashed, dot-dashed and short-dashed curves, respectively.  
In addition, the gray straight line shows the resolution limit of the simulations due to the shot-noise contribution. 
The values of $\zeta$ calculated in terms of the simulations results assuming $k_{max}=50\kMpc$ are given in column 2 of Table~\ref{tabzeta} for $z=0$, 
$1$ and $2$, showing a good agreement among numerical results and fits/models. But, clearly, the behavior of the DM  field at scales well below the resolution presently achieved by N-body simulations is  dominating the flux. 
In order to estimate the error from an extrapolation of the simulations results we can make a very simple assumption inspired by direct inspection of the data: $\Delta_{NL}(k)$ is a non-decreasing function of $k$, but its second derivative is negative  at sufficiently large scales. Hence, we assume that the true $\Delta_{NL}(k)$ is bracketed by the 
two following cases
\bea
\Delta_{NL}^{min}(k)\!\!\!\!&=&\!\!\!\!\Delta_{NL}(k^\star)\,,\:\:{\rm for}\:k>k^\star\,,\label{extrap0}\\
\Delta_{NL}^{max}(k)\!\!\!\!&=&\!\!\!\!\Delta_{NL}(k^\star)\!+\!\Delta_{NL}'(k^\star)(k\!-\!k^\star)\,,\:{\rm for}\:k>k^\star\,,\label{extrap1}
\end{eqnarray}
where $k^\star$ denotes the scale beyond which the spectrum cannot be trusted anymore because of numerical resolution effects, and the prime stands for derivative with respect to $k$.  For illustration, in columns 7 and 12 of Table~\ref{tabzeta} we report in the format  ``min''/``max'' the results of such extrapolations done beyond $k^\star=\{29, 16,9.8\}\,\kMpc$ for $z=0,1,2$, respectively. The $k^\star$'s have been fixed here as the values where the expected shot-noise component reaches 1\% of $\Delta_{NL}(k^\star,z)$. The extrapolation is then performed correcting the data for shot-noise.
The columns 7 and 12 of Table~\ref{tabzeta} show how $\zeta(z)$ depends in fact on the $k_{max}$ value. Note that, although the functional form of the fitting functions of Eq.~(\ref{extrap0}) and Eq.~(\ref{extrap1}) is heuristical, several different approaches to modeling nonlinear gravitational clustering suggest a flattening of $\Delta(k)$ at large $k$, \ie $\Delta'(k)\to 0$  at large $k$ \citep[see \eg][]{PadmanabhanRay2006, AnguloWhite2010}, giving some confidence that  Eq.~(\ref{extrap0}) and Eq.~(\ref{extrap1}) represent extreme cases.

The results in Table~\ref{tabzeta} illustrate the following points:

\vspace{-0.1cm}
\begin{itemize}

\item At least up to $k_{\rm max}=10^4\,h/$Mpc (middle set of columns), the comparison between the the \texttt{halofit} formula and the extended HM is reasonable, with  \texttt{halofit} expectations falling in the middle of the \gio\, predictions at low $z$, while slightly below at higher $z$; notice however that  HM predictions are typically overshooting simulations at high $z$ \citep[see \eg][]{KlypinTrujilloGomezPrimack2011}. Additionally, the simple extrapolations of Eq.s~(\ref{extrap0}) and~(\ref{extrap1}) nicely bracket these model expectations and confirms a posteriori that they can be seen as reasonable error estimates on the signal.

\item The last columns represent our best guess for a conservative error range and best estimate of $\zeta(z)$, assuming the representative particle physics input $k_{max}=10^6\kMpc$. Based on the MS, the value $\zeta(0)\approx 10^{7.5}$ appears as a conservative upper limit, and $\zeta(0)\approx 10^{4.3}$ a conservative lower-limit. On the other hand, a similar preliminary analysis for the result of the Millennium-II simulation extracted from Fig. 6 in~\citep{BoylanKolchinEtal2009} yields  {\it one order of magnitude reduction} in the extrapolation error, as shown in column 13,  thanks to the higher resolution (we estimated $k^\star=\{212,120,84\}\,\kMpc$ for $z=0,1,2$). Less conservative estimates of $k^\star$ or a more careful treatment may further reduce the error.
\end{itemize}
\vspace{-0.1cm}

Of course, our results do not include effects unaccounted for in the pure cold DM models considered here, nor  additional particle physics and cosmology assumptions. Without the need to discuss any uncertainty on auxiliary variables such as concentration, mass function, substructure, etc. we reproduced the range of uncertainty computed with traditional methods in configuration space \citep[see \eg][Fig.~1]{AbdoEtal2010} with the MS data, while improving by one order of magnitude over it when using MS-II data.   Also the decreasing functional shape of $\zeta(z)$ is in rough agreement with existing computations. Obviously, this is a simpler and more direct method than what currently employed in the literature and it makes obvious that no information on separate ingredients of the HM can be inferred by EDMF.

\vspace{-0.2cm}
%%%%%%%%%%%%%
\section{Discussion and conclusions}\label{conclusions}
%%%%%%%%%%%%%
In the context of WIMP models for DM, the extragalactic flux from the residual annihilation of these particles represents one possible handle for the discovery of the particle physics nature of DM, as well as perhaps the only tool to extract cosmological informations from very small scales. 
In this article we revisited the issues involved in the computation of the main cosmological ingredient for this signal, the so-called flux multiplier $\zeta(z)$. Our results suggest that a direct computation of $\zeta(z)$ in terms of nonlinear power spectrum presents several advantages with respect to traditional HM computations in configuration space. In short, it allows one to\vspace{-0.1cm}
\begin{itemize}
\item isolate the essential physical quantity which the signal depends on, merely $\Delta_{NL}(k,z)$, rather than requiring to fit several theoretical functions, which in the HM parameterize
halo-related observables which are useful in many instances, but irrelevant {\it for the present application};
\item  bypass the problems linked to ill-defined concepts or different algorithms used by different research groups and publications to isolate each auxiliary variable: the PS constitute in principle a {\em complete} description of nonlinear matter clustering relevant for $\zeta(z)$ (of course, bearing in mind the problem of knowing the UV behaviour, see discussion in Sec.~\ref{pscomputation} and footnote 2);  
\item ease the extrapolation problem: one is reduced to find a plausible extrapolation for a single function and for ``only'' about 4 decades in $k$ (under the conventional assumptions on the UV behaviour of $\Delta_{NL}(k,z)$ referred to above) making the estimate much less prone to cumulative systematic effects of a wrong choice of {\it several} fitting functions. Even when working with the HM, one {\em should} gauge the plausibility of different extrapolations by comparison with the PS yielded by numerical simulations; to put it simply {\em such information is available and should not be discarded};
\item ease the task of estimating a credible  error budget for $\zeta(z)$: simple extrapolations of $\Delta_{NL}(k,z)$, suggested by inspection of its behavior in the resolved regime, proved to be as spread as (for MS) and significantly better (for MS-II) than the existing range of HM estimates. This promising results will of course require more extensive analyses and inspection of a wider set of simulations (as well as for effects like the ones of baryons, neutrinos, etc.), which we leave for future work.  
\end{itemize}
\vspace{-0.1cm}

Also, one might envisage to develop further insights along related directions. One example would be to assess the extent to which one may combine information from simulations available for structures at different scales, \eg high-resolution cosmological simulation such as Millennium-II and galactic-scales simulations such as Aquarius~\citep{SpringelEtal2008} and Via Lactea~\citep{KuhlenEtal2008}. Note that  the cosmological PS is already commonly obtained from simulations by separating ``large scales'' and ``smaller scales''~\citep{JenkinsEtal1998, SpringelEtal2005}: by choosing the small cells so that they contain statistically the same initial (linear) power, each one contributes about the same to the final power, which allows for a relatively sparse sampling. Similarly, 
as long as one carries out high-resolution simulations of ``representative'' cells
it might be possible to extend the range of the nonlinear PS which can be directly extracted from simulations. 

Additionally, one might extend the reasoning to compute the anisotropy in the EDMF: 
since its 2PCF is related to the variance of the DM power spectrum, computing the ``$C_l$'s'' only requires additionally the ``squeezed'' four point correlation function.
Although some early attempts to derive the signal purely on the basis of simulation results exist~\citep{CuocoEtal2007}, it would be interesting to combine some recent theoretical insights with modern simulations.  In fact, compared with the average EDMF, a larger fraction of the measurable anisotropy signal should follow from the weakly nonlinear scales (see \eg\cite{AndoKomatsu2006}.)

Finally, although the HM has proven quite flexible and effective in describing most scales of interest for cosmology in the last decade, there are some theoretical arguments why it might be less and less suitable to describe the large$-k$ and high-$z$ behavior of $\Delta_{NL}(k,z)$. As argued for example in~\cite{ElahiEtal2009}, at smaller and smaller scales haloes do not fully virialize before being accreted, with the boundary of (sub)structures becoming increasingly ill defined. Extrapolations based on mass functions computed at galactic scales to the bottom of the CDM hierarchy are thus questionable. It would be important to decouple the expected growing uncertainties and inadequacies of a specific (albeit till now successful) model from the actual dependence of the signal of interest here from the underlying cosmology, which is more neatly encoded in a direct computation in terms of the power spectrum or two-point correlation function.

%%%%%%%%%%%%%%%%%%%%%%%%%%%%%%%%%%%%%%%%%%%%%%%%%%%%%%%%%%%%%%%%%%%%%%%%%%%%%%%%%%%%%%%%%%%%%
\medskip\noindent{\bf Acknowledgments:}
We thank C. Giocoli,  R. Sheth, and J. Zavala for useful correspondence. MG acknowledge the support from the Fondazione Cariparo Excellence Grant `LHC and Cosmology', PS and GZ  the support of ``Mati{\`e}re noire: Des observations cosmologiques aux mod{\`e}les microscopiques" from INP of CNRS within the `PEPS Projects - Physique Th{\'e}orique et ses Interfaces', and ES the Marie Curie IEF program. This work was partly carried out at CERN within the TH-Institute DMUH'11 (18-29 July 2011).
\vspace{-0.8cm}
\bibliography{Bibliography}

%%%%%%%%%%%%%%%%%%%%%%%%%%%%%%%%%%%%%%%%%%%%%%%%%%%%%%%%%%%%%%%%%%%%%
\end{document}